\documentclass[aps,pra,twocolumn,groupedaddress]{revtex4-2}

\usepackage{graphicx}
\usepackage{amsfonts}
\usepackage{amsmath}
\usepackage{amssymb}
\usepackage{braket}
\usepackage{mathtools}
\usepackage{siunitx}
\usepackage{accents}
\usepackage[utf8]{inputenc}
\usepackage{circledsteps}
\usepackage{kbordermatrix}
\usepackage{xcolor}
\usepackage{enumerate}
\usepackage{mathtools}
\usepackage{cases}
\usepackage{bm}
\usepackage{empheq}
\usepackage{braket}
\usepackage{comment}
\usepackage[makeroom]{cancel}

\usepackage{stackengine}
\stackMath
\newcommand\tsup[2][2]{%
	\def\useanchorwidth{T}%
	\ifnum#1>1%
	\stackon[-.5pt]{\tsup[\numexpr#1-1\relax]{#2}}{\scriptscriptstyle\sim}%
	\else%
	\stackon[.5pt]{#2}{\scriptscriptstyle\sim}%
	\fi%
}

\newcommand{\Lim}[1]{\raisebox{0.5ex}{\scalebox{0.8}{$\displaystyle \lim_{#1}\;$}}}

\begin{document}

\title{Maximally Nonlinear and Nonconservative Quantum Circuits}

\author{Matteo Mariantoni}
\email[Corresponding author: ]{matteo.mariantoni@uwaterloo.ca}
\affiliation{Institute for Quantum Computing, University of Waterloo, 200 
University Avenue West, Waterloo, Ontario N2L 3G1, Canada}
\affiliation{Department of Physics and Astronomy, University of Waterloo, 200 
University Avenue West, Waterloo, Ontario N2L 3G1, Canada}

\date{\today}

\begin{abstract}
In this article, we introduce an algorithmic method to find the conservative energy and non-conservative power of a large class of maximally nonlinear electric circuits (including Josephson tunnel junctions, coherent quantum phase slips, and superconducting loops), based on the incidence matrix of the circuits' digraph. We consider two-port linear circuits with mostly holonomic constraints provided by either Maxwell-Kirchhoff's current rules or Maxwell-Kirchhoff's voltage rules. The circuit's independent variables, generally a superset of the degrees of freedom, are obtained from the solution space of Maxwell-Kirchhoff's current or voltage rules. The method does not require to find any Lagrangian. Instead, the circuit's classical or quantum Hamiltonian is obtained from the energy of the reactive (i.e., conservative) circuit elements by means of transformations complementary to Hamilton's equations. Dissipation (loss) is accounted for by using the Rayleigh dissipation function and defining generalized Poisson brackets\textemdash Poisson-Rayleigh brackets. Fluctuations (noise) are added as voltage or current sources characterized by bath modes. Non-conservative elements (e.g., noisy resistors) are included \textit{ab initio} using the incidence-matrix method, without needing to treat them as separate elements. Finally, we show that in order to form a complete set of canonical coordinates, auxiliary (which could be parasitic in certain cases) circuit elements are required to find the Hamiltonian of circuits with an incomplete set of generalized velocities. In particular, we introduce two methods to eliminate the coordinates associated with the auxiliary elements by either Hamiltonian reduction or equation-of-motion reduction. We use auxiliary circuit elements to treat a maximally nonlinear circuit comprising simultaneously both a Josephson junction and a quantum phase slip.
\end{abstract}


\maketitle

\section{Introduction}

Electric circuit theories analogous to classical Hamiltonian mechanics and quantum mechanics have been developed by many authors in the past century. The first article we are aware of is by D.A.~Wells in 1938~\cite{Wells:1938}, where linear circuits are treated. Over the next several decades, a large body of work culminated with the theory of nonlinear circuits by B.M.~Maschke \textit{et al.} in Ref.~\cite{Maschke:1995}. More recently, the development of quantum computers based on superconducting circuits has renewed the interest in this topic; the works by Burkard \textit{et al.} in Ref.~\cite{Burkard:2004} and by Vool and Devoret in Ref.~\cite{Vool:2017} explore similar approaches, although the former follows a very rigorous method and the latter a more practical one. Other works include the quantum network theory by Yurke and Denker~\cite{Yurke:1984} as well as the Foster representation method of Russer and Russer~\cite{Russer:2011}. Additionally, it is worth mentioning the seminal works on noisy resistors by J.B.~Johnson~\cite{Johnson:1928} and H.~Nyquist~\cite{Nyquist:1928}, which are still very relevant for treating circuit dissipation and fluctuations, as well as the so-called input-output theory by Gardiner and Collet~\cite{Gardiner:1985}.

\section{From Energy to Hamiltonian without Lagrangian}

\begin{figure}[b]
	\centering
	\includegraphics{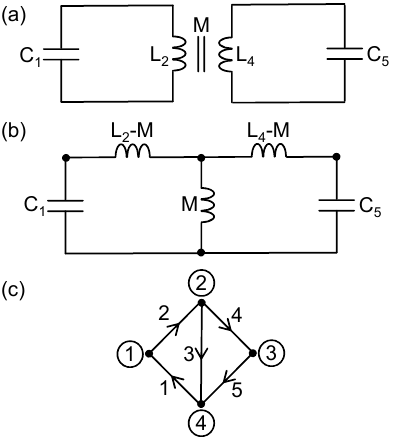}
\caption{(a) Two inductively coupled resonators. (b) Connected circuit. (c) Digraph.}
	\label{fig:circuit1}
\end{figure}

The first objective of this work is to introduce an algorithmic method to find the classical Hamiltonian of a conservative electric circuit, without needing any Lagrangian. We begin by analyzing the electric circuit shown in Fig.~\ref{fig:circuit1}~(a) and, then, generalize the method to the study of more general two-port linear circuits with holonomic constraints. Throughout the article, we indicate a potential difference (or voltage) as~$v$; a current as~$\imath$ (without dot, not to get confused with time derivatives); a flux as~$\phi$; a charge as~$q$.

The physical circuit in Fig.~\ref{fig:circuit1}~(a) is comprised of two inductively coupled resonators. Each resonator is realized as the parallel connection of a capacitor with self-capacitance~$C_1$ or $C_5$ and an inductor with self-inductance~$L_2$ or $L_4$; the two resonators are coupled by means of an inductor with mutual inductance~$M$. Excluding any parasitic capacitance, the physical inductively coupled circuit is unconnected. By means of a well-known artifice~\cite{Chua:1987}, we can draw an equivalent connected circuit as diagrammed in Fig.~\ref{fig:circuit1}~(b).

Figure~\ref{fig:circuit1}~(c) illustrates the digraph~\cite{Chua:1987} associated with the connected circuit in Fig.~\ref{fig:circuit1}~(b). The digraph is comprised of~$n = 4$ nodes, $\mathcal{N} = \{ \Circled{1}, \Circled{2}, \Circled{3}, \Circled{4} \}$, as well as~$b = 5$ oriented branches, $\mathcal{B} = \{ 1, 2, \ldots, 5 \}$; we set~$\Circled{4} \rightarrow \textrm{datum}$, i.e., to a reference node experimentally realized by earthing or grounding the circuit. The sign convention for the oriented branches with respect to a node is that any branch entering the node is given the value~$-1$ and any branch exiting it the value~$1$; if a branch does not enter or exit the node, it is given the value~$0$.

The branch currents (as well as charges), the branch voltages (as well as fluxes), and the node-to-datum voltages are represented by the vectors
\begin{subequations}
	\begin{empheq}[left=\empheqlbrace]{alignat=2}
\vec{\imath}^{\, \textrm{T}}
& = \left[ \imath_1 \, \imath_2 \, \imath_3 \, \imath_4 \, \imath_5 \right]
& \,\,\, ( \vphantom{\left[ q_1 \, q_2 \, q_3 \, q_4 \, q_5 \right]}
\vec{q}^{\, \textrm{T}}
& =	
\left[ q_1 \, q_2 \, q_3 \, q_4 \, q_5 \right] )
	\label{subeq:ivec}
\\[2.5mm]
\vec{v}^{\, \textrm{T}}
& = \left[ v_1 \, v_2 \, v_3 \, v_4 \, v_5 \right]
& \,\,\, ( \vphantom{\left[ \phi_1 \, \phi_2 \, \phi_3 \, \phi_4 \, \phi_5 \right]}
\vec{\phi}^{\, \textrm{T}}
& =	
\left[ \phi_1 \, \phi_2 \, \phi_3 \, \phi_4 \, \phi_5 \right] )
	\label{subeq:vvec}
\\[2.5mm]
\vec{e}^{\, \textrm{T}}
& = \left[ e_1 \, e_2 \, e_3 \, e_4 \right]
	\enskip ,
	\label{subeq:evec}
	\end{empheq}
\end{subequations}
where~$[\cdot]^{\textrm{T}}$ indicates the transpose of a vector or matrix~$[\cdot]$.

Following the digraph sign convention outlined above, the incidence matrix (see, e.g., Ref.~\cite{Chua:1987}) for the digraph of Fig.~\ref{fig:circuit1} is
\renewcommand{\kbldelim}{(}
\renewcommand{\kbrdelim}{)}
\begin{equation}
\mathbf{A}_{a}
	= \kbordermatrix{
	& \phantom{-} 1 & \phantom{-} 2 & \phantom{-} 3 & \phantom{-} 4 & \phantom{-} 5 \\
	\Circled{1} \enskip \rightarrow & -1 & \phantom{-} 1 & \phantom{-} 0 & \phantom{-} 0 & \phantom{-} 0 \\
	\Circled{2} \enskip \rightarrow & \phantom{-} 0 & - 1 & \phantom{-} 1 & \phantom{-} 1 & \phantom{-} 0 \\
	\Circled{3} \enskip \rightarrow & \phantom{-} 0 & \phantom{-} 0 & \phantom{-} 0 & - 1 & \phantom{-} 1 \\
	\Circled{4} \enskip \rightarrow & \phantom{-} 1 & \phantom{-} 0 & - 1 & \phantom{-} 0 & - 1
					}
\enskip .
	\label{eq:incidence:matrix}
\end{equation}
The reduced incidence matrix~$\mathbf{A}$ is found by striking out the row in~$\mathbf{A}_{a}$ associated with the datum, resulting in
\begin{equation}
\mathbf{A} =
	\begin{bmatrix}
		- 1 & \phantom{-} 1 & \phantom{-} 0 & \phantom{-} 0 & \phantom{-} 0 \\
		\phantom{-} 0 & -1 & \phantom{-} 1 & \phantom{-} 1 & \phantom{-} 0 \\
		\phantom{-} 0 & \phantom{-} 0 & \phantom{-} 0 & - 1 & \phantom{-} 1
	\end{bmatrix}
\enskip .
	\label{eq:reduced:incidence:matrix}
\end{equation}

In matrix form, Maxwell-Kirchhoff's current laws (or better rules)~(for simplicity, KCLs here) and Maxwell-Kirchhoff's voltage laws~(KVLs) are~$\mathbf{A} \, \vec{\imath} = 0$ and $\vec{v} = \mathbf{A}^{\! \textrm{T}} \, \vec{e}$, respectively.

Among all possible vectors~$\vec{\imath}$, the set of branch current vectors satisfying~KCL is called the \emph{KCL solution subspace}~$K_{\imath}$; the dimension of~$K_{\imath}$ is~$D_{\imath} = b-n+1$. Similarly, we can define the \emph{KVL solution subspace}~$K_v$, the dimension of which is~$D_v = n-1$ (see Ref.~\cite{Chua:1987}). The subspace associated with the independent variables~(IVs) (which, in general, are a superset of the degrees of freedom) of the circuit has dimension~$D = \min( D_{\imath} , D_v )$.

For the circuit of Fig.~\ref{fig:circuit1}~(b), $D = \min( D_{\imath}=5-4+1 , D_v=4-1 ) = 2$. Thus, $D = D_{\imath} = 2$ and the circuit should be solved using KCL: This is a KCL circuit. The KCL relation at node~$\Circled{2}$ gives~$\imath_3 = \imath_2 - \imath_4$. This constraint sets the branch currents associated with the two~IVs to be~$\imath_2$ and $\imath_4$.

The voltage branch equation for a generic inductor with current~$\imath_L = \accentset{\bullet}{q}_L$ and inductance~$L$ is $v_L = \accentset{\bullet}{\phi}_L = L \accentset{\bullet}{\imath}_L$, where~$\phi_L$ is the inductor's flux. The instantaneous inductive energy of the circuit is obtained by integrating the inductive power from an initial time~$t' = 0$ to a generic time~$t' = t$,
\begin{widetext}
\begin{eqnarray}
\mathcal{E}_L (t)
& = & \int_{0}^{t} dt' \, \mathcal{P}_L = \int_{0}^{t} dt' \, ( v_2 \imath_2 + v_3 \imath_3 + v_4 \imath_4 )
\nonumber\\
& = & \int_{0}^{t} dt' \dfrac{d}{dt'} \left[ \dfrac{1}{2} (L_2 - M) \, \accentset{\bullet}{q}_2^2 (t') + \dfrac{1}{2} M (\accentset{\bullet}{q}_2 - \, \accentset{\bullet}{q}_4)^2 (t') + \dfrac{1}{2} (L_4 - M) \, \accentset{\bullet}{q}_4^2 (t') \right]
\enskip .
	\label{eq:EL24:from:PL24}
\end{eqnarray}
\end{widetext}
Hereafter, unless otherwise specified, we set all initial conditions to zero [e.g., $\accentset{\bullet}{q}_2 (0) = \accentset{\bullet}{q}_4 (0)$] and hide the time dependence of all variables. We obtain,
\begin{equation}
\mathcal{E}_L = \dfrac{1}{2} L_2 \, \accentset{\bullet}{q}_2^2 - M \, \accentset{\bullet}{q}_2 \accentset{\bullet}{q}_4 + \dfrac{1}{2} L_4 \, \accentset{\bullet}{q}_4^2
\enskip .
	\label{eq:EL24}
\end{equation}

We still have to use the two remaining KCL relations at nodes~$\Circled{1}$ and $\Circled{3}$, yielding~$\accentset{\bullet}{q}_1 = \accentset{\bullet}{q}_2$ and $\accentset{\bullet}{q}_4 = \accentset{\bullet}{q}_5$. These two constraints are equivalent to
\begin{subequations}
	\begin{empheq}[left=\empheqlbrace]{align}
		q_1 & = q_2 + \tsup[1]{q}
\label{subeq:q1}
		\\[2mm]
		q_5 & = q_4 + \tsup[2]{q}
		\enskip ,
\label{subeq:q5}
	\end{empheq}
\end{subequations}
where~$\tsup[1]{q}$ and $\tsup[2]{q}$ are constant charge offsets. It is easy to show that these offsets (which would correspond to DC voltages in the circuit's equations of motion) must be further constrained due to Tellegen's theorem~\cite{Chua:1987}.

Similarly to~$\mathcal{E}_L$, the instantaneous capacitive energy of the circuit reads
\begin{eqnarray}
\mathcal{E}_C
& = & \int_{0}^{t} dt' \, \mathcal{P}_C = \int_{0}^{t} dt' \dfrac{d}{dt'} \left[ \dfrac{1}{2} C_1 \, \accentset{\bullet}{\phi}_1^2 (t') + \dfrac{1}{2} C_5 \, \accentset{\bullet}{\phi}_5^2 (t') \right ]
\nonumber\\
& = & \dfrac{1}{2} C_1 \, \accentset{\bullet}{\phi}_1^2 + \dfrac{1}{2} C_5 \, \accentset{\bullet}{\phi}_5^2
\enskip .
	\label{eq:EC15:from:PC15}
\end{eqnarray}

The total circuit's energy~$\mathcal{E} = \mathcal{E}_L + \mathcal{E}_C$ encodes the exact same information as the Hamiltonian; however, $\mathcal{E}$ is written in terms of the two distinguished sets of generalized velocities $\{ \accentset{\bullet}{q}_2 , \accentset{\bullet}{q}_4 \}$ and $\{ \accentset{\bullet}{\phi}_1 , \accentset{\bullet}{\phi}_5 \}$, whereas we expect the Hamiltonian to be written in terms of one set of canonical coordinates~$\{ \phi_2 , \phi_4 ; q_2 , q_4 \}$. In order to obtain the Hamiltonian from~$\mathcal{E}$, we need to solve the two systems of equations
\begin{subequations}
	\begin{empheq}[left=\empheqlbrace]{align}
	& \left\{
	\begin{array}{ll}
		\dfrac{\partial}{\partial \accentset{\bullet}{\phi}_1} \mathcal{E}
		& = q_1 \\
		[5.0mm]
		\dfrac{\partial}{\partial \accentset{\bullet}{\phi}_5} \mathcal{E}
		& = q_5
	\end{array}
	\right.
\label{subeq:Hamilton:eqs:dual:q15}
\\[2.5mm]
	& \left\{
	\begin{array}{ll}
		\dfrac{\partial}{\partial \accentset{\bullet}{q}_2} \mathcal{E}
		& = - \phi_2 \\
		[5.0mm]
		\dfrac{\partial}{\partial \accentset{\bullet}{q}_4} \mathcal{E}
		& = - \phi_4
	\end{array}
	\right.
	\enskip .
\label{subeq:Hamilton:eqs:dual:phi24}
	\end{empheq}
\end{subequations}
These equations are analogous to Hamilton's equations~\cite{Cercignani:1976}. This explains the \emph{minus sign} on the right-hand side of Eqs.~(\ref{subeq:Hamilton:eqs:dual:phi24}). Nevertheless, we prefer to invert these equations to obtain the circuit's Hamiltonian for three main reasons: Firstly, and chiefly, by substituting Eqs.~(\ref{subeq:q1}) and (\ref{subeq:q5}) into the right-hand side of Eqs.~(\ref{subeq:Hamilton:eqs:dual:q15}), we are able to write all subsequent equations only as a function of the~IV set~$\{ \phi_2 , \phi_4 ; q_2 , q_4 \}$ (solving the apparent issue of using~$q_1$ and $q_5$ as independent variables); secondly, when considering Josephson tunnel junctions~(JTJs) and quantum phase slips~(QPSs) (see below), it is highly desirable (practically necessary) to use charge and flux variables instead of voltages and currents; lastly, the Hamiltonian is more standard than the energy~$\mathcal{E}$ and, thus, better suited for quantization.

In matrix form, the system of Eqs.~(\ref{subeq:Hamilton:eqs:dual:phi24}) becomes
\begin{equation}
\begin{bmatrix*}[c]
L_2 & M \\
M   & L_4
\end{bmatrix*}
\begin{bmatrix*}[c]
\accentset{\bullet}{q}_2 \\
\accentset{\bullet}{q}_4
\end{bmatrix*}
\triangleq
\mathbf{L} \, \vec{\accentset{\bullet}{q}}_{\imath}
=
- \vec{\phi}_{\imath}
\triangleq
\begin{bmatrix*}[c]
- \phi_2 \\
- \phi_4
\end{bmatrix*}
\enskip ,
	\label{eq:L:matrix24}
\end{equation}
where~$\mathbf{L}$ is the inductance matrix. Similarly,
\begin{equation}
\begin{bmatrix*}[c]
C_1 & 0 \\
0   & C_5
\end{bmatrix*}
\begin{bmatrix*}[c]
\accentset{\bullet}{\phi}_1 \\
\accentset{\bullet}{\phi}_5
\end{bmatrix*}
\triangleq
\mathbf{C} \, \vec{\accentset{\bullet}{\bar{\phi}}}_{\imath}
=
- \vec{q}_{\imath}
\triangleq
\begin{bmatrix*}[c]
(q_2 + \tsup[1]{q}) \\
(q_4 + \tsup[2]{q})
\end{bmatrix*}
\enskip ,
	\label{eq:C:matrix24}
\end{equation}
where~$\mathbf{C}$ is the capacitance matrix. By solving the system of Eqs.~(\ref{eq:L:matrix24}) for~$\accentset{\bullet}{q}_2$ and $\accentset{\bullet}{q}_4$ and the system of Eqs.~(\ref{eq:C:matrix24}) for~$\accentset{\bullet}{\phi}_1$ and $\accentset{\bullet}{\phi}_5$ and substituting the results into Eqs.~(\ref{eq:EL24}) and (\ref{eq:EC15:from:PC15}) and summing, we find the circuit's Hamiltonian
\begin{eqnarray}
\mathcal{H}
& = & \dfrac{1}{\textrm{det} \mathbf{L}} \left( \dfrac{L_4}{2} \phi_2^2 + \phi_2 \, M \, \phi_4 + \dfrac{L_2}{2} \phi_4^2 \right)
\nonumber\\
& + & \dfrac{\left( q_2 + \tsup[1]{q} \right)^2}{2 C_1} + \dfrac{\left( q_4 + \tsup[2]{q} \right)^2}{2 C_5}
\enskip .
	\label{eq:H24}
\end{eqnarray}
From~$\mathcal{H}$ we can find the circuit's equations of motion, as well as the quantized Hamiltonian if needed.

In general, assume the sets~$\{ \phi_k , \accentset{\bullet}{\phi}_k \}$ and $\{ q_k , \accentset{\bullet}{q}_k \}$ represent two distinguished sets of generalized coordinates and their corresponding time derivatives (i.e., generalized velocities), with~$k \in \mathbb{N}$; the variables~$q_k$ are also the conjugate momenta of the generalized coordinates~$\phi_k$. Therefore, $\{ \phi_k , q_k \}$ is a set of canonical coordinates. However, depending whether we are studying a KVL or KCL circuit, the total circuit energy usually reads
\begin{equation}
\mathcal{E_\text{{tot}}} = \mathcal{E_\text{{tot}}} ( \phi_i , \accentset{\bullet}{\phi}_i ; q_j , \accentset{\bullet}{q}_j )
\enskip ,
	\label{eq:E:tot:mixed}
\end{equation}
where the actual~IV variable set is either~$\{ \phi_i , \accentset{\bullet}{\phi}_i \}$ for a KVL circuit or $\{ q_j , \accentset{\bullet}{q}_j \}$ for a KCL circuit, with~$i \neq j$; the variable set is determined by the primary set of constraints associated with KVLs or KCLs and, possibly, any necessary reduction constraints associated with the KCLs for nodes connected with auxiliary circuit elements when the primary constraints are KVLs or the KVLs for loops containing auxiliary circuit elements when the primary constraints are KCLs (see below). The transformations required to find the circuit's Hamiltonian are then
\begin{subequations}
	\begin{empheq}[left=\empheqlbrace]{align}
		\dfrac{\partial}{\partial \accentset{\bullet}{\phi}_i} \mathcal{E_{\text{tot}}} ( \phi_i , \accentset{\bullet}{\phi}_i ; q_j , \accentset{\bullet}{q}_j )
		& = q_i
\label{eq:generalized:qi}
\\[2mm]
		\dfrac{\partial}{\partial \accentset{\bullet}{q}_j} \mathcal{E_{\text{tot}}} ( \phi_i , \accentset{\bullet}{\phi}_i ; q_j , \accentset{\bullet}{q}_j )
		& = - \phi_j
\enskip .
\label{eq:generalized:phij}
	\end{empheq}
\end{subequations}
The relationship between the~$j$ and $i$ indexes for a KVL circuit and between the~$i$ and $j$ indexes for a KCL circuit are found by integrating in time the KVL or KCL relations, respectively. It is rather easy to write this formally for a KCL circuit,
\begin{equation}
A_{n i} \, q_i + \sum_{\substack{j = 1 \\ j \neq i}}^{b - 1} A_{n j} \, q_j = \tilde{q}_i
\enskip ,
	\label{eq:integrated:KCL}
\end{equation}
where~$A_{n b}$ is the element for node~$n$ and branch~$b$ of~$\mathbf{A}$ and $\tilde{q}_i$ is a constant of integration. A similar system of equations applies for a KVL circuit.

\section{Non-Conservative Linear Circuits}

Now that we are acquainted with the method, we use it to solve a \emph{non-conservative circuit}, i.e., a circuit with noisy resistors. Irrespective of the circuit's resistive nature, to further extend our method we purposely choose a circuit that results in an incomplete set of canonical coordinates. We conjecture that any circuit of this class requires the inclusion of an \emph{auxiliary circuit element} (in real applications, this could be a parasitic element; for the purposes of this theory, auxiliary circuits are ``mathematical tools'' that we need initially and, then, eliminate by reduction) in correspondence to the missing coordinate to complete the set. We then show a reduction that allows us to eliminate the coordinate(s) associated with the auxiliary element.

\begin{figure}[t!]
	\centering
	\includegraphics{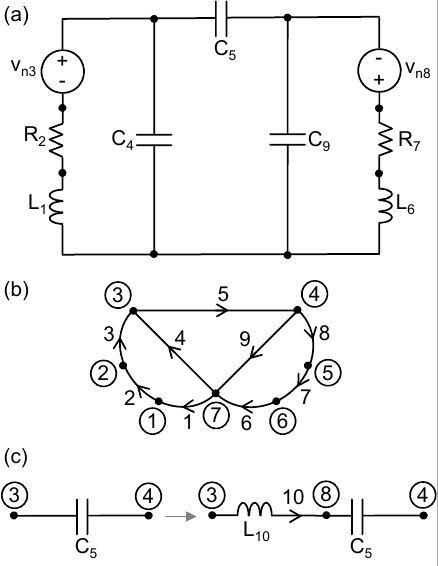}
	\caption{(a) Capacitively coupled resonators with noisy resistors. (b) Digraph. (c) Auxiliary circuit element.}
	\label{fig:circuit2}
\end{figure}

Figure~\ref{fig:circuit2}~(a) shows the physical circuit of a pair of resonators consisting of the parallel connection of inductors with inductance~$L_1$ or $L_6$ and capacitors with capacitance~$C_4$ or $C_9$; the two resonators are coupled by means of a capacitor with capacitance~$C_5$. The inductors are assumed to be connected in series with noisy resistors; following Nyquist~\cite{Nyquist:1928}, each noisy resistor is modelled as a noiseless resistor with resistance~$R_2$ for $L_1$ and $R_7$ for $L_6$ in series with a noise source (Helmholtz-Th\'{e}venin equivalent circuit) of root mean square voltage~$v_{\textrm{n} 3} = \sqrt{4 k_{\textrm{B}} T R_2 \Delta f}$ and $v_{\textrm{n} 8} = \sqrt{4 k_{\textrm{B}} T R_7 \Delta f}$ (where~$k_{\textrm{B}}$ is the Boltzmann constant, $T$ the circuit's thermodynamic temperature, and $\Delta f$ a given frequency bandwidth assuming additive white Gaussian noise). To avoid a cumbersome calculation, without a significant loss of generality, we assume all the capacitors to be ideal circuit elements, i.e., with negligible resistance. In general, \emph{any inductor should be accompanied by a series noisy resistor and any capacitor by a parallel noisy resistor}.

Figure~\ref{fig:circuit2}~(b) illustrates the digraph for the circuit in Fig.~\ref{fig:circuit2}~(a). In this case, $n = 7$ and $b = 9$ and, thus, $D_{\imath} = 3$, $D_v = 6$, and $D = D_{\imath} = 3$. Even though it is tempting to use~KVL due to the capacitive network embedded in the circuit, $D = D_{\imath}$ prescribes the usage of~KCL. It is worth noting that not using~KVL does not result in a loss of information, since the entire topology of the circuit is already built in the incidence matrix~$\mathbf{A}$ used in~KCLs. Note that, when neglecting the noisy resistors in the circuit of Fig.~\ref{fig:circuit2}~(a), $D = D_v = 2$. In that case, our method prescribes to use KVL, clearly showing the duality in the usage of KCLs and KVLs depending on the circuit topology.

The circuit of Fig.~\ref{fig:circuit2} is characterized by a total power~$\mathcal{P} = \mathcal{P}^{(\textrm{c})} + \mathcal{P}^{(\textrm{nc})}$, where~$\mathcal{P}^{(\textrm{c})}$ and $\mathcal{P}^{(\textrm{nc})}$ are the \emph{conservative} and \emph{non-conservative} power, respectively. The powers dissipated in the noiseless resistors add up to form~$\mathcal{P}^{(\textrm{nc})}$; all the other circuit elements are conservative and their power can be trivially integrated in time.

Using KCL at nodes~$\Circled{1}$ and $\Circled{2}$ as well as $\Circled{5}$ and $\Circled{6}$ and noting that the power associated with each voltage noise, $- v_{\textrm{n} 3} (d q_1/dt)$ and $- v_{\textrm{n} 8} (dq_6/dt)$ (the minus sign is due to the standard convention to invert a voltage source sign compared to its branch current), can be readily integrated in time to obtain the corresponding energies, $- v_{\textrm{n} 3} q_1$ and $- v_{\textrm{n} 8} q_6$, the total conservative energy of the circuit is
\begin{widetext}
\begin{equation}
\mathcal{E}
=
\dfrac{1}{2} L_1 \, \accentset{\bullet}{q}_1^2 - v_{\textrm{n} 3} q_1 + \dfrac{1}{2} C_4 \accentset{\bullet}{\phi}_4^2 + \dfrac{1}{2} C_5 \accentset{\bullet}{\phi}_5^2 + \dfrac{1}{2} C_9 \accentset{\bullet}{\phi}_9^2 + \dfrac{1}{2} L_6 \accentset{\bullet}{q}_6^2 - v_{\textrm{n} 8} q_6
\enskip .
	\label{eq:E:circuit2}
\end{equation}
\end{widetext}

Considering that~$D = 3$ and we are using~KCL, we expect to have three independent branch currents in~$\mathcal{E}$. However, we only have two independent currents in Eq.~(\ref{eq:E:circuit2}). We must add an \emph{auxiliary circuit element} to find the third independent current. In order to choose where to add the auxiliary element and what type of element to add, we first perform the transformation of Eq.~(\ref{eq:generalized:qi}) for the three capacitive elements in Eq.~(\ref{eq:E:circuit2}) and integrate in time KCL at nodes~$\Circled{3}$ and $\Circled{4}$ as prescribed by Eq.~(\ref{eq:integrated:KCL}); we find
\begin{subequations}
	\begin{empheq}[left=\empheqlbrace]{align}
	C_4 \accentset{\bullet}{\phi}_4
	& =
	q_4 = q_5 - q_1 + \tsup[1]{q}
\label{eq:C4:phi4dot}
\\[2mm]
	C_5 \accentset{\bullet}{\phi}_5
	& =
	q_5
	\label{eq:C5:phi5dot}
\\[2mm]
	C_9 \accentset{\bullet}{\phi}_9
	& =
	q_9 = q_5 - q_6 + \tsup[2]{q}
\enskip .
	\label{eq:C9:phi9dot}
	\end{empheq}
\end{subequations}
Clearly, the missing current is~$\accentset{\bullet}{q}_5$. It is well known that any capacitor is characterized by a series parasitic inductor. Therefore, we add an auxiliary inductor with inductance~$L_{10}$ in series with~$C_5$, as shown in Fig.~\ref{fig:circuit2}~(c). It is worth noting that the addition of~$L_{10}$ preserves~$D = D_{\imath} = 3$.

Note that if we have to include an auxiliary element for an inductor, we would add a parallel capacitor; for a resistive wire, a series inductor (or vice versa); for a resistor, a parallel capacitor, with this parallel connection in series with an inductor; for a transformer as in Fig.~\ref{fig:circuit1}~(a), a pair of capacitors connecting the left and right circuits above and below~\cite{Chua:1987}; etc. However, in some cases the auxiliary circuit element does not have to necessarily correspond to a parasitic element: It can simply be a ``mathematical'' auxiliary element that we then attempt to eliminate by means of a reduction.

By means of~KCL at node~$\Circled{8}$, we find~$q_5 = q_{10} + \bar{q}$ (where~$\bar{q}$ is a constant charge). We can thus rewrite Eqs.~(\ref{eq:C4:phi4dot}), (\ref{eq:C5:phi5dot}), and (\ref{eq:C9:phi9dot}) as
\begin{subequations}
	\begin{empheq}[left=\empheqlbrace]{align}
	C_4 \accentset{\bullet}{\phi}_4
	& = q_{10} - q_1 + \bar{\tsup[1]{q}}
\label{eq:C4:phi4dot:auxiliary}
\\[2mm]
	C_5 \accentset{\bullet}{\phi}_5
	& = q_{10} + \bar{q}
	\label{eq:C5:phi5dot:auxiliary}
\\[2mm]
	C_9 \accentset{\bullet}{\phi}_9
	& = q_{10} - q_6 + \bar{\tsup[2]{q}}
\enskip ,
	\label{eq:C9:phi9dot:auxiliary}
	\end{empheq}
\end{subequations}
where~$\bar{\tsup[1]{q}} = \tsup[1]{q} + \bar{q}$ and $\bar{\tsup[2]{q}} = \tsup[2]{q} + \bar{q}$.

Due to the addition of~$L_{10}$, we must include an extra energy term~$L_{10} \accentset{\bullet}{q}_{10}^2 / 2$ to~$\mathcal{E}$. The transformations of Eq.~(\ref{eq:generalized:phij}) for the three inductive elements read
\begin{subequations}
	\begin{empheq}[left=\empheqlbrace]{align}
	L_1 \accentset{\bullet}{q}_1
	& = - \phi_1
\label{eq:L1:q1dot:auxiliary}
\\[2mm]
	L_{10} \accentset{\bullet}{q}_{10}
	& = - \phi_{10}
	\label{eq:L10:q10dot:auxiliary}
\\[2mm]
	L_6 \accentset{\bullet}{q}_6
	& = - \phi_6
\enskip .
	\label{eq:L6:q6dot:auxiliary}
	\end{empheq}
\end{subequations}

By inverting Eqs.~(\ref{eq:C4:phi4dot:auxiliary}), (\ref{eq:C5:phi5dot:auxiliary}), (\ref{eq:C9:phi9dot:auxiliary}), (\ref{eq:L1:q1dot:auxiliary}), (\ref{eq:L10:q10dot:auxiliary}), and (\ref{eq:L6:q6dot:auxiliary}) to find the generalized velocities as a function of the complete canonical-coordinate set~$\{ \phi_1 , \phi_{10} , \phi_6 ; q_1 , q_{10} , q_6 \}$ and substituting the results into~$\mathcal{E}$, we obtain the circuit's Hamiltonian
\vspace{10.0mm}
\begin{widetext}
\begin{equation}
\mathcal{H}
=
\dfrac{\phi_1^2}{2 L_1} - v_{\textrm{n} 3} q_1 + \dfrac{(q_{10} - q_1 + \bar{\tsup[1]{q}})^2}{2 C_4} + \dfrac{\phi_{10}^2}{2 L_{10}} + \dfrac{(q_{10} + \bar{q})^2}{2 C_5} + \dfrac{(q_{10} - q_6 + \bar{\tsup[2]{q}})^2}{2 C_9} - v_{\textrm{n} 8} q_6 + \dfrac{\phi_6^2}{2 L_6}
\enskip .
	\label{eq:H:circuit2}
\end{equation}
\end{widetext}
Since both~$q_1$ and $q_6$ interact with~$q_{10}$, as expected there is an effective interaction between~$q_1$ and $q_6$ (i.e., between the two resonators).

While it is true that a physical parasitic inductor~$L_{10}$ leads to a circuit with three canonical coordinates, it is unsettling not to be able to reduce the original circuit of Fig.~\ref{fig:circuit2}~(a) to just two canonical coordinates. In fact, the original circuit represents two simple capacitively-coupled (dissipative) harmonic oscillators. Thus, we would expect only two degrees of freedom and therefore two canonical coordinates.

We can attempt to eliminate the additional coordinate associated with the auxiliary harmonic oscillator~$L_{10} C_5$ by either obtaining the three KVLs from Eq.~(\ref{eq:H:circuit2}) and using the KVL for the central loop (formed by the three capacitors) or by writing down this KVL by inspection. Since the final result is the same, we choose the latter method as it requires fewer calculations. In order to select the proper extra KVL, we consider the ciruit of ~\ref{fig:circuit2}~(a) with the addition of the auxiliary circuit element, as shown in Fig.~~\ref{fig:circuit2}~(c).

The KVL required for the reduction is the one involving the auxiliary element,
\begin{equation}
v_4 + v_5 + v_{10} + v_9 = 0 \enskip .
\end{equation}
Inserting the constitutive relations for the various elements and taking the limit for~$L_{10} \rightarrow 0^{+}$ (short-circuit condition), this constraint becomes
\begin{eqnarray}
\textrm{KVL} \, 10
& = &
\lim_{L_{10} \rightarrow 0^{+}} \left( \dfrac{q_4}{C_4} + \dfrac{q_{10}}{C_5} + \dfrac{q_9}{C_9} + L_{10} \accentset{\bullet\bullet}{q}_{10} \right) \nonumber\\
& = & \dfrac{q_4}{C_4} + \dfrac{q_{10}}{C_5} + \dfrac{q_9}{C_9} = 0 \enskip .
\end{eqnarray}
This is an holonomic constraint that allows us to reduce the canonical coordinates from three to two by eliminating~$q_{10}$:
\begin{equation}
q_{10} = \dfrac{q_1}{\tilde{C}_4} + \dfrac{q_6}{\tilde{C}_9} \enskip ,
	\label{eq:q10}
\end{equation}
where~$\tilde{C}_4 = \alpha C_4$, $\tilde{C}_9 = \alpha C_9$, and $\alpha = (1/C_4 + 1/C_5 + 1/C_9)$. Notably, this holonomic reduction is equivalent to a~$Y-\Delta$ transformation for the~$C_4-C_5-C_9$ subnetwork.

We now insert the constraint of Eq.~(\ref{eq:q10}) into Eq.~(\ref{eq:H:circuit2}), set the inductive energy of~$L_{10}$ to zero (short-circuit limit), set~$v_{\textrm{n}3} = v_{\textrm{n}8} = 0$ for simplicity, and assume~$\bar{\tsup[1]{q}} = \bar{q} = \bar{\tsup[2]{q}} = 0$, we find the reduced Hamiltonian
\begin{widetext}
\begin{equation}
\mathcal{\widetilde{H}}
=
\dfrac{\phi_1^2}{2 L_1} + \dfrac{\left( \dfrac{q_1}{\tilde{C}_4} + \dfrac{q_6}{\tilde{C}_9} - q_1 \right)^2}{2 C_4} + \dfrac{\left( \dfrac{q_1}{\tilde{C}_4} + \dfrac{q_6}{\tilde{C}_9} \right)^2}{2 C_5} + \dfrac{\left( \dfrac{q_1}{\tilde{C}_4} + \dfrac{q_6}{\tilde{C}_9} - q_6 \right)^2}{2 C_9} + \dfrac{\phi_6^2}{2 L_6}
\enskip .
	\label{eq:Htilde:circuit2}
\end{equation}
\end{widetext}

The total non-conservative power cannot be integrated in time using the usual stratagems used for all conservative circuit elements. Hence, we describe all non-conservative elements by means of their power instead of their energy. For the circuit in Fig.~\ref{fig:circuit2}, from~KCL at nodes~$\Circled{1}$ and $\Circled{6}$ and Ohm's law, $v_2 = R_2 \imath_1$ and $v_7 = R_7 \imath_6$, the total non-conservative power reads
\begin{equation}
\mathcal{P}^{(\textrm{nc})} = v_2 \imath_2 + v_7 \imath_7 = R_2 \accentset{\bullet}{q}_1^2 + R_7 \accentset{\bullet}{q}_6^2
\enskip ,
	\label{eq:P:nc}
\end{equation}
which confirms Joule-Lenz law (Joule's first law) for each resistor.

\subsection{Poisson-Rayleigh Brackets}

From classical mechanics~\cite{Cercignani:1976} and Tellegen's theorem
\begin{equation}
\dfrac{d}{dt} \mathcal{H} + 2 \mathcal{D} = \dfrac{d}{dt} \mathcal{H} + \mathcal{P}^{(\textrm{nc})} = 0
\enskip ,
	\label{eq:zero:power:mechanics}
\end{equation}
where~$\mathcal{D} = \mathcal{P}^{(\textrm{nc})} / 2$ is the Rayleigh dissipation function. Assuming ohmic resistors (i.e., isotropic, linear, and homogeneous resistive conductors),
\begin{equation}
\mathcal{D} = \sum_{\ell = 1}^{N^{\parallel}_{\textrm{r}}} \dfrac{1}{2} G_{\ell} \, ( \accentset{\bullet}{\phi}_{\ell} )^2 + \sum_{m = 1}^{N^{-}_{\textrm{r}}} \dfrac{1}{2} R_m \, ( \accentset{\bullet}{q}_m )^2
\enskip ,
	\label{eq:D}
\end{equation}
where~$N^{\parallel}_{\textrm{r}} , N^{-}_{\textrm{r}} \in \mathbb{N}$ are, respectively, the total number of resistors with conductance~$G_{\ell}$ and voltage~$\accentset{\bullet}{\phi}_{\ell}$ and resistors with resistance~$R_m$ and current~$\accentset{\bullet}{q}_m$. The voltage~$\accentset{\bullet}{\phi}_{\ell} \neq 0$ when using KVLs as primary constraints, whereas the current~$\accentset{\bullet}{q}_m \neq 0$ when KCLs are the primary constraints. Typically, both~$\accentset{\bullet}{\phi}_{\ell}$ and $\accentset{\bullet}{q}_m$ can be written in terms of the canonical coordinates identified when deriving the Hamiltonian. In certain cases, the resistor voltages or currents may require additional generalized velocities outside the set of those associated with the canonical coordinates. In these instances, it is necessary to perform a resistive reduction as elucidated in one example below.

For the circuit of Fig.~\ref{fig:circuit2}~(a), we have
\begin{equation}
\mathcal{D} = \dfrac{1}{2} R_2 \accentset{\bullet}{q}_1^2 + \dfrac{1}{2} R_7 \accentset{\bullet}{q}_6^2 \enskip .
	\label{eq:D:circuit2}
\end{equation}
As expected, only terms with~$\accentset{\bullet}{q}_k$ are present since we are using KCLs as primary constraints. In this case, the dissipative subnetwork can be written as a function of the canonical coordinates used in the reduced Hamiltonian of Eq.~(\ref{eq:Htilde:circuit2}).

In order to write the circuit's equations of motion in presence of dissipation, we extend the concept of Poisson brackets~\cite{Cercignani:1976} by defining a new tool, the \emph{generalized Poisson brackets}. For any pair of canonical coordinates~$( \phi_k , q_k )$, they are
\begin{subequations}
	\begin{empheq}[left=\empheqlbrace]{align}
	\accentset{\bullet}{q}_k
	& = \{ ( \mathcal{H} , \mathcal{D} ) , ( q_k , \accentset{\bullet}{q}_k ) \} \nonumber\\
	& = \sum_k \left( \dfrac{\partial \mathcal{H}}{\partial q_k} \dfrac{\partial q_k}{\partial \phi_k} - \dfrac{\partial \mathcal{H}}{\partial \phi_k} \dfrac{\partial q_k}{\partial q_k} \right) \nonumber\\
	& + \sum_k \left( \dfrac{\partial \mathcal{D}}{\partial \accentset{\bullet}{q}_k} \dfrac{\partial \accentset{\bullet}{q}_k}{\partial \accentset{\bullet}{\phi}_k} - \dfrac{\partial \mathcal{D}}{\partial \accentset{\bullet}{\phi}_k} \dfrac{\partial \accentset{\bullet}{q}_k}{\partial \accentset{\bullet}{q}_k} \right)
\label{eq:genPoisson:qkdot}
\\[2mm]
	\accentset{\bullet}{\phi}_k
	& = \{ ( \mathcal{H} , \mathcal{D} ) , ( \phi_k , \accentset{\bullet}{\phi}_k ) \} \nonumber\\
		& = \sum_k \left( \dfrac{\partial \mathcal{H}}{\partial q_k} \dfrac{\partial \phi_k}{\partial \phi_k} - \dfrac{\partial \mathcal{H}}{\partial \phi_k} \dfrac{\partial \phi_k}{\partial q_k} \right) \nonumber\\
		& + \sum_k \left( \dfrac{\partial \mathcal{D}}{\partial \accentset{\bullet}{q}_k} \dfrac{\partial \accentset{\bullet}{\phi}_k}{\partial \accentset{\bullet}{\phi}_k} - \dfrac{\partial \mathcal{D}}{\partial \accentset{\bullet}{\phi}_k} \dfrac{\partial \accentset{\bullet}{\phi}_k}{\partial \accentset{\bullet}{q}_k} \right)
	\enskip .
\label{eq:genPoisson:phikdot}
	\end{empheq}
\end{subequations}

After performing some algebra, the generalized Poisson brackets for the reduced Hamiltonian of Eq.~(\ref{eq:Htilde:circuit2}) and for the dissipation function of Eq.~(\ref{eq:D:circuit2}) allow us to find, e.g., the dual KVL associated with the loop containing~$L_1$, $R_2$, and $C_4$ [obtained by setting~$k = 1$ in Eqs.~(\ref{eq:genPoisson:qkdot}) and (\ref{eq:genPoisson:phikdot})]
\begin{subequations}
	\begin{empheq}[left=\empheqlbrace]{align}
	\accentset{\bullet}{q}_1
	& = - \dfrac{\phi_1}{L_1}
\label{eq:Poisson:q1dot}
\\[2mm]
	\accentset{\bullet}{\phi}_1
	& = \left[ \dfrac{1}{C_4} \left( \dfrac{1}{\tilde{C}_4} - 1 \right)^{\! 2} + \dfrac{1}{C_5} \dfrac{1}{\tilde{C}_4^2} + \dfrac{1}{C_9} \dfrac{1}{\tilde{C}_4^2} \right] q_1 \nonumber\\
	& + \bigg[ \dfrac{1}{C_4} \left( \dfrac{1}{\tilde{C}_4} - 1 \right) \dfrac{1}{\tilde{C_9}} + \dfrac{1}{C_5} \dfrac{1}{\tilde{C}_4 \tilde{C}_9} \nonumber\\
	& + \dfrac{1}{C_9} \dfrac{1}{\tilde{C}_4} \left( \dfrac{1}{\tilde{C}_9} - 1 \right) \bigg] q_6 + R_2 \accentset{\bullet}{q}_1
	\enskip .
\label{eq:Poisson:phi1dot}
	\end{empheq}
\end{subequations}
Solving Eq.~(\ref{eq:Poisson:q1dot}) for~$\phi_1$, deriving the result with respect to~$t$ and substituting into Eq.~(\ref{eq:Poisson:phi1dot}), we finally find
\begin{equation}
L_1 \accentset{\bullet\bullet}{q}_1 + R_2 \accentset{\bullet}{q}_1 - \dfrac{1}{C_4} \left( \dfrac{1}{\tilde{C}_4} - 1 \right) q_1 - \dfrac{1}{C_4 \tilde{C}_9} q_6 = 0
	\enskip .
\label{eq:second:order:ODE:1}
\end{equation}
This is the same equation we would obtain by applying directly KVL to this loop and using the necessary KCL constraints and branches constitutive relations~\footnote{Note that it is straightforward to include the contribution from~$v_{\textrm{n} 3}$ in Eq.~(\ref{eq:second:order:ODE:1})}. Similar equations are found for the other IVs, forming a system.

\begin{figure}[t!]
	\centering
	\includegraphics{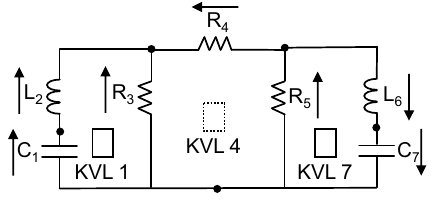}
\caption{Highly dissipative circuit. Each branch current is oriented in the same direction as the corresponding voltage. For simplicity, and without loosing generality, we neglect the fluctuation terms associated with each resistor. The solid-line boxes indicate the two KVLs dual to the primary KCLs. The dotted-line box shows the reduction KVL constraint.}
	\label{fig:circuit4}
\end{figure}

\subsection{Dissipative Holonomic Reductions and Nonholonomic Reductions}

We now consider the highly dissipative circuit displayed in Fig.~\ref{fig:circuit4}. For this circuit, $D = D_{\imath} = 3 < D_v = 4$ and, thus, KCLs are the primary circuit laws, whereas KVLs the dual ones. By setting up KCLs, it is straightforward to obtain the three independent circuit currents, which are the currents on $C_1$, $R_4$, and $C_7$, $\accentset{\bullet}{q}_1$, $\accentset{\bullet}{q}_4$, and $\accentset{\bullet}{q}_7$.

Following the procedure outlined in the previous examples, the circuit Hamiltonian reads
\begin{equation}
\mathcal{H}
=
\dfrac{q_1^2}{2 C_1} + \dfrac{\phi_1^2}{2 L_2} + \dfrac{q_7^2}{2 C_7} + \dfrac{\phi_7^2}{2 L_6}
\enskip .
	\label{eq:H:circuit4}
\end{equation}
Similarly, the dissipation function reads
\begin{equation}
\mathcal{D} = \dfrac{1}{2} R_3 \, ( \accentset{\bullet}{q}_1 + \accentset{\bullet}{q}_4 )^2 + \dfrac{1}{2} R_4 \, \accentset{\bullet}{q}_4^2 + \dfrac{1}{2} R_5 \, ( \accentset{\bullet}{q}_4 + \accentset{\bullet}{q}_7 )^2
\enskip .
	\label{eq:D:circuit4}
\end{equation}

The circuit is characterized by an incomplete set of canonical coordinates due to the presence of the independent generalized velocity~$\accentset{\bullet}{q}_4$ in~$\mathcal{D}$ and the absence of~$q_4$ and $\phi_4$ in~$\mathcal{H}$. In this case, however, it is not necessary to introduce any auxiliary circuit element because~$\mathcal{H}$ itself is ``complete'' with respect to the set of canonical coordinates~$\{ \phi_1 , \phi_7 ; q_1 , q_7 \}$.

In presence of an auxiliary circuit, we would have attempted to perform a reduction by adding to the set of KCLs the dual KVL associated with a loop containing the auxiliary element. This time, instead, we perform the reduction by considering the KVL associated with the loop of three resistive branches, $\accentset{\bullet}{\phi}_3 - \accentset{\bullet}{\phi}_4 - \accentset{\bullet}{\phi}_5 = 0$, which is the loop containing the ``extra'' current~$\accentset{\bullet}{q}_4$. Inserting the constitutive relation for each resistor, we obtain the reduction constraint
\begin{equation}
\accentset{\bullet}{q}_4 = -\tilde{R}_3 \accentset{\bullet}{q}_1 - \tilde{R}_5 \accentset{\bullet}{q}_7 \enskip ,
	\label{eq:q4dot}
\end{equation}
where~$\tilde{R}_3 = R_3/\alpha$, $\tilde{R}_5 = R_5/\alpha$, and $\alpha = R_3 + R_4 + R_5$.

Inserting Eq.~(\ref{eq:q4dot}) into Eq.~(\ref{eq:D:circuit4}) we find the reduced dissipation function~$\mathcal{\widetilde{D}}$, which depends only on~$\accentset{\bullet}{q}_1$ and $\accentset{\bullet}{q}_7$. The generalized Poission brackets for KVL~1 (the loop associated with KVL~1 is indicated in Fig.~\ref{fig:circuit4}) then read
\begin{subequations}
	\begin{empheq}[left=\empheqlbrace]{align}
	\accentset{\bullet}{q}_1
	& = \{ ( \mathcal{H} , \mathcal{\widetilde{D}} ) , ( q_1 , \accentset{\bullet}{q}_1 ) \} = - \dfrac{\phi_1}{L_2}
\label{eq:gen:Poisson:KVL1:q}
\\[2mm]
	\accentset{\bullet}{\phi}_1
	& = \{ ( \mathcal{H} , \mathcal{\widetilde{D}} ) , ( \phi_1 , \accentset{\bullet}{\phi}_1 ) \} \nonumber\\
	& = \dfrac{q_1}{C_1} + R_3 ( 1 - \tilde{R}_3 ) \accentset{\bullet}{q}_1 - R_5 \tilde{R}_3 \accentset{\bullet}{q}_7 \enskip .
\label{eq:gen:Poisson:KVL1:phi}
	\end{empheq}
\end{subequations}
The resulting equation of motion for the loop associated with KVL~1 is thus
\begin{equation}
\dfrac{q_1}{C_1} + L_2 \accentset{\bullet\bullet}{q}_1 + R_3 ( 1 - \tilde{R}_3 ) \accentset{\bullet}{q}_1 - R_5 \tilde{R}_3 \accentset{\bullet}{q}_7 = 0 \enskip ,
\end{equation}
which is the same equation as obtained by circuit direct inspection. We leave to find the equation of motion for the loop associated with KVL~7 as an exercise (in fact, this is a good exercise to verify all the signs are correct).

\begin{figure}[t!]
	\centering
	\includegraphics{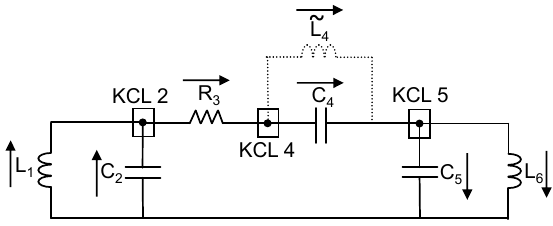}
\caption{Circuit with nonholonomic reduction.}
	\label{fig:circuit5}
\end{figure}

We now turn to the circuit of Fig.~\ref{fig:circuit5}. Before adding the auxiliary circuit element~$\tilde{L}_4$, the circuit is characterized by~$D_{\imath} = D_v = 3$. In such a degenerate case, we can use either KVLs or KCLs as primary constraints. In this case, we elect to use KVLs and find
\begin{equation}
\mathcal{H} = \dfrac{\phi_2^2}{2 L_1} + \dfrac{q_2^2}{2 C_2} + \dfrac{\phi_4^2}{2 \tilde{L}_4} + \dfrac{q_4^2}{2 C_4} + \dfrac{\phi_5^2}{2 L_6} + \dfrac{q_5^2}{2 C_5}
\end{equation}
and
\begin{equation}
\mathcal{D} = \dfrac{1}{2} \dfrac{\accentset{\bullet}{\phi}_2^2 + \accentset{\bullet}{\phi}_4^2 + \accentset{\bullet}{\phi}_5^2}{R_3} \enskip .
\end{equation}

We could attempt to reduce the number of canonical coordinates to two by imposing KCL reduction constraints at nodes~$4$ and $5$, KCL~$4$ and $5$, and then impose the condition~$\lim \tilde{L}_4 \rightarrow + \infty$ (open circuit condition). However, these constraints form a nonholonomic system and, thus, cannot be used to perform the reduction.

In this case, we keep the auxiliary circuit elements in the unreduced Hamiltonian and find the equations of motion associated with all three (or as many as present in other similar circuits) canonical coordinates and impose the $\lim \tilde{L}_4 \rightarrow + \infty$ for these equations at the end. This allows us to eliminate the auxiliary quantity~$\tilde{L}_4$ from the equations of motion. This is highly desirable as we do not want to necessarily quantify~$\tilde{L}_4$ to solve the problem.

Using our generalized Poisson equations, the equation of motion, e.g., for node~$5$ reads
\begin{equation}
\lim_{\tilde{L}_4 \rightarrow +\infty} \left( C_4 \accentset{\bullet\bullet}{\phi}_4 + \dfrac{\phi_4}{\tilde{L}_4} + \dfrac{\accentset{\bullet}{\phi}_2 + \accentset{\bullet}{\phi}_4 + \accentset{\bullet}{\phi}_5}{R_3} \right) = 0 \enskip .
\end{equation}

\section{Nonlinear and Nonconservative Circuit with JTJ: Quantization and Fluctuations}

\begin{figure}[t]
	\centering
	\includegraphics{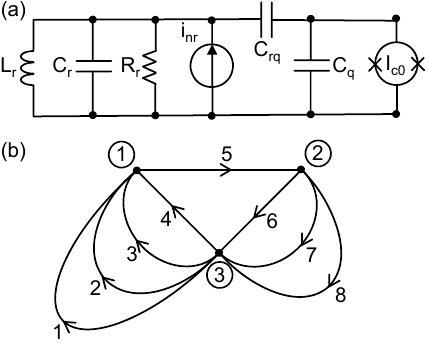}
\caption{(a) Non-conservative resonator capacitively coupled to transmon qubit. (b) Digraph.}
	\label{fig:circuit3}
\end{figure}

We now consider a circuit consisting of a non-conservative~$L_{\textrm{r}} C_{\textrm{r}} R_{\textrm{r}}$ parallel resonator coupled by means of a capacitor with capacitance~$C_{\textrm{rq}}$ to (for simplicity) a conservative flux-tunable transmon qubit~\cite{Barends:2013}. The transmon qubit is comprised of a capacitor with capacitance~$C_{\textrm{q}}$ connected in parallel with a~SQUID~\cite{Barone:1982}. The SQUID is realized as the parallel connection of two~JTJs with (for simplicity) equal critical current~$I_{\textrm{c} 0}$~\cite{Barone:1982}. The circuit and its digraph are illustrated in Fig.~\ref{fig:circuit3}. We first set~$R_{\textrm{r}} \rightarrow +\infty$ (open circuit) and, thus, $\imath_{\textrm{nr}} = 0$.

This nonlinear circuit is easy to treat due to the fact that the~JTJs are in parallel with a capacitor: This is equivalent to having a phase-controlled current source in parallel with a capacitor. If the junctions were in series with the capacitor, the circuit would have been harder but can be solved employing auxiliary circuit elements.

When neglecting the quasiparticle and ``cos~$\varphi$'' terms, the constitutive relation (or first Josephson relation) of the JTJ at branch~$u = 7 , 8$ is
\begin{equation}
\imath ( \phi_u )
=
I_{\textrm{c} 0} \, \sin ( k_{\text{J}} \phi_u )
=
I_{\textrm{c} 0} \, \sin ( \varphi_u ) \enskip ,
\label{eq:Josephson:equation}
\end{equation}
where~$k_{\textrm{J}} = 2 \pi / \Phi_0$ is the non-normalized Josephson constant and $\varphi$ the gauge-invariant phase difference across the junction; $\Phi_0 = h / ( 2 e )$ is the superconducting magnetic flux quantum ($h$ is the Planck's constant and $e$ the electron charge).

From the equivalent of Faraday-Henry-Neumann's law for JTJs (or second Josephson relation), the voltage across each junction is given by
\begin{equation}
v_u
=
\accentset{\bullet}{\phi}_u
=
\dfrac{\accentset{\bullet}{\varphi}_u}{k_{\text{J}}}
\enskip .
	\label{eq:vu:phiudot}
\end{equation}

From the circuit digraph we find~$D = D_v = 2$. Using KVL,
\begin{widetext}
\begin{equation}
\mathcal{E}
=
\dfrac{1}{2} L_{\textrm{r}} \, \accentset{\bullet}{q}_1^2 + \dfrac{1}{2} C_{\textrm{r}} \accentset{\bullet}{\phi}_2^2 + \dfrac{1}{2} C_{\textrm{rq}} \left( -\accentset{\bullet}{\phi}_2 - \accentset{\bullet}{\phi}_6 \right)^2 + \dfrac{1}{2} C_{\textrm{q}} \accentset{\bullet}{\phi}_6^2 + \mathcal{E}_{\textrm{J}}
\enskip ,
	\label{eq:E:circuit3}
\end{equation}
\end{widetext}
where~$\mathcal{E}_{\textrm{J}}$ is the Josephson energy.

In this case, the parallel conditions~$v_6 = v_7$ and $v_7 = v_8$ become
\begin{subequations}
	\begin{empheq}[left=\empheqlbrace]{align}
		\phi_6 - \phi_7 & = \tsup[2]{\phi}
		\label{subeq:phi6:phi7}
\\[2mm]
		\phi_7 - \phi_8 & = k_{\textrm{q}} \Phi_0
		\enskip ,
\label{subeq:flux:quantization}
	\end{empheq}
\end{subequations}
with~$k_{\textrm{q}} \in \mathbb{Z}$. Equation~(\ref{subeq:flux:quantization}) is a special case of the condition following from Faraday-Henry-Neumann's law and it is called \emph{flux quantization condition}; this condition must be used in presence of any superconducting loop~\cite{Tinkham:1996}.

From simple algebra and trigonometric identities,
\begin{eqnarray}
\mathcal{E}_{\textrm{J}}
& = &
\int_0^t dt' \, \accentset{\bullet}{\phi}_6 \,\, I_{\textrm{c} 0} \, [ \sin ( k_{\textrm{J}} \phi_7 ) + \sin ( k_{\textrm{J}} \phi_8 ) ]
\nonumber\\
& = &
- 2 E_{\textrm{J} 0} \, \cos [ k_{\textrm{J}} ( \phi_6 - \tsup[2]{\phi} ) ] + K_{\textrm{J}}
\enskip ,
	\label{eq:EJ}
\end{eqnarray}
where~$E_{\textrm{J} 0} = I_{\textrm{c} 0} / k_{\textrm{J}}$ and $K_{\textrm{SQUID}}$ is a constant of integration that hereafter we set to zero, $K_{\textrm{SQUID}} = 0$. Thus, the pair of junctions in the SQUID can be treated as a single effective junction with twice the Josephson energy of each junction. Flux tunability can be included by coupling inductively a current source to the SQUID loop.

Applying the transformations of Eqs.~(\ref{eq:generalized:qi}) and (\ref{eq:generalized:phij}) to Eq.~(\ref{eq:E:circuit3}) and using the condition from KVL, $\phi_1 = \phi_2 + \tsup[1]{\phi}$, the circuit Hamiltonian reads
\begin{widetext}
\begin{equation}
\mathcal{H}_{\textrm{rq}}
=
\dfrac{(\phi_2 + \tsup[1]{\phi})^2}{2 L_{\textrm{r}}}
+
\dfrac{1}{\textrm{det} \, \mathbf{C}} \left[ \dfrac{C_{\textrm{q}} + C_{\textrm{rq}}}{2} q_2^2 + q_2 \, C_{\textrm{rq}} \, q_6 + \dfrac{C_{\textrm{r}} + C_{\textrm{rq}}}{2} q_6^2 \right]
- 2 E_{\textrm{J} 0} \, \cos [ k_{\textrm{J}} ( \phi_6 - \tsup[2]{\phi} ) ]
\enskip ,
	\label{H:circuit3:classical}
\end{equation}
\end{widetext}
with
\begin{equation}
\mathbf{C}
	=
	\begin{bmatrix*}[c]
	(C_{\textrm{r}} + C_{\textrm{rq}}) & C_{\textrm{rq}} \\[1.0mm]
	C_{\textrm{rq}} & (C_{\textrm{q}} + C_{\textrm{rq}}) \\[1.0mm]
	\end{bmatrix*}
	\enskip .
\label{eq:C:circuit3}
\end{equation}

Following a standard quantization procedure~\cite{Loudon:2000}, the classical canonical coordinates are promoted to quantum-mechanical operators as
\begin{subequations}
	\begin{empheq}[left=\empheqlbrace]{align}
	\left( \phi_2 , \phi_6 \right)
	& \rightarrow
	\left( \hat{\phi}_2 , \hat{\phi}_6 \right)
\label{subeq:vecphi:quantized}
	\\
	\left( q_2 , q_6 \right)
	& \rightarrow
	\left( \hat{q}_2 , \hat{q}_6 \right) = \left( -\jmath \hbar \dfrac{\partial}{\partial \hat{\phi}_2} , -\jmath \hbar \dfrac{\partial}{\partial \hat{\phi}_6} \right)
\label{subeq:vecq:quantized}
\enskip ,
	\end{empheq}
\end{subequations}
where~$\jmath^2 = -1$ and $\hbar = h / (2 \pi)$.

We now set~$C_{\textrm{rq}} = 0$, $R_{\textrm{r}} \neq 0$ and finite, and consider only the resonator circuit. For this circuit, $D = D_v = 1$ and from KVL and Eqs.~(\ref{eq:generalized:qi}) and (\ref{eq:generalized:phij}) we find
\begin{equation}
\mathcal{H}_{\textrm{r}}^{'} = \dfrac{( \phi_2 + \tsup[1]{\phi} )^2}{2 L_{\textrm{r}}} + \dfrac{q_2^2}{2 C_{\textrm{r}}} + \phi_2 \, \imath_{\textrm{nr}}
	\label{eq:H:resonator:classical}
\enskip .
\end{equation}
Defining~$\mathcal{H}_{\textrm{nr}} = \phi_2 \, \imath_{\textrm{nr}}$, we can write~$\mathcal{H}_{\textrm{r}}^{'} = \mathcal{H}_{\textrm{r}} + \mathcal{H}_{\textrm{nr}}$.

In this case, $\mathcal{D} = \accentset{\bullet}{\phi}_2^2 / ( 2 R_{\textrm{r}} )$. The generalized Poisson bracket term~$( \partial / \partial \accentset{\bullet}{\phi}_2 ) \mathcal{D} = - \accentset{\bullet}{q}_3$ allows us to find the classical equation of motion
\begin{equation}
\accentset{\bullet}{q}_2 = \{ \mathcal{H}_{\textrm{r}}^{'} , q_2 \} - \dfrac{\accentset{\bullet}{\phi}_2}{R_{\textrm{r}}}
\enskip .
	\label{eq:q2dot:classical}
\end{equation}

Using the standard creation and annihilation operators~$\hat{a}^{\dagger}$ and $\hat{a}$, $\mathcal{H}_{\textrm{r}} = h f_{\textrm{r}} \left( \hat{a}^{\dagger} \hat{a} + 1 / 2 \right)$, where~$f_{\textrm{r}} = 1 / ( 2 \pi \sqrt{L_{\textrm{r}} C_{\textrm{r}}} )$. Quantizing the Poisson brackets, Eq.~(\ref{eq:q2dot:classical}) leads to the quantum Langevin equation
\begin{equation}
\accentset{\bullet}{\hat{q}}_2 = \dfrac{1}{\jmath \hbar} \left\{ [ \mathcal{\hat{H}}_{\! \textrm{r}} , \hat{q}_2 ] + [ \mathcal{\hat{H}}_{\! \textrm{nr}} , \hat{q}_2 ] \right\} - \dfrac{\accentset{\bullet}{\hat{\phi}}_2}{R_{\textrm{r}}}
\enskip .
	\label{eq:q2dot:quantum}
\end{equation}

From the quantum version of Nyquist theorem and assuming a bosonic noise bath~$\hat{b}$ with small bandwidth~$\Delta f$ around~$f_{\textrm{r}}$, $\hat{\imath}_{\textrm{nr}} = I_{\textrm{n} 0} ( \hat{b}^{\dagger} + \hat{b} )$, with~$I_{\textrm{n} 0} = \sqrt{2 h f_{\textrm{r}} \Delta f / R_{\textrm{r}}}$ (see, e.g., Ref.~\cite{Mariantoni:2010}). Since~$\hat{\phi}_2 = \phi_0 ( \hat{a}^{\dagger} + \hat{a} )$ and $\hat{q}_2 = q_0 \jmath ( \hat{a}^{\dagger} - \hat{a} )$, with~$\phi_0 = \sqrt{L_{\textrm{r}} h f_{\textrm{r}} / 2}$ and $q_0 = \sqrt{C_{\textrm{r}} h f_{\textrm{r}} / 2}$, it is easy to prove that
\begin{equation}
\dfrac{1}{\jmath \hbar} [ \mathcal{\hat{H}}_{\textrm{nr}} , \hat{q}_2 ] = \dfrac{2 \phi_0 I_{\textrm{n} 0} q_0}{\hbar} ( \hat{b}^{\dagger} + \hat{b} )
\enskip ,
	\label{eq:inr:quantized}
\end{equation}
as expected.

\section{Maximally Nonlinear Circuits: Hybrid Formalism}

\begin{figure*}[t!]
	\centering
	\includegraphics[width=0.561\textwidth]{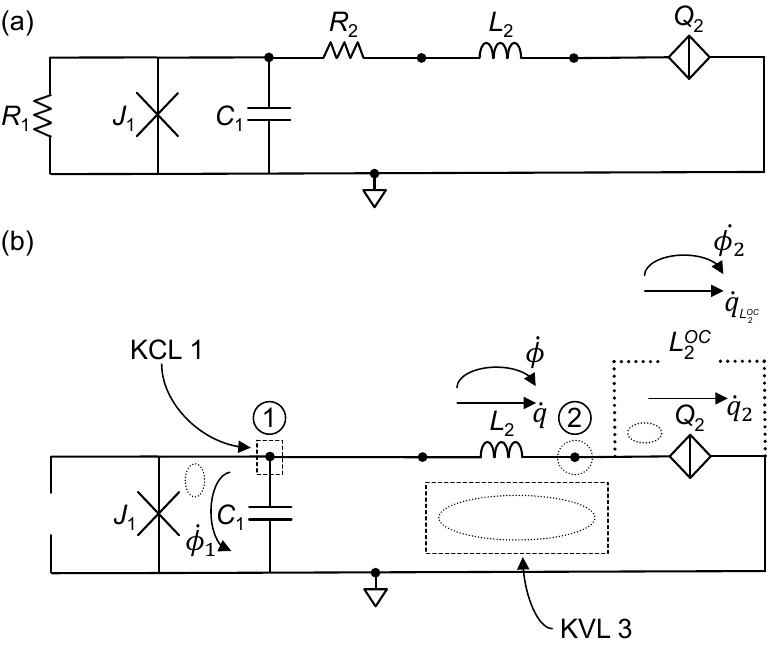}
	\caption{Maximally nonlinear circuit. (a) A~JTJ named~$J_1$ with parallel resistance~$R_1$ and capacitance~$C_1$ is connected to a~QPS named~$Q_2$ with series resistance~$R_2$ and inductance~$L_2$. (b) Conservative circuit associated with the circuit in~(a). $L_2^{\text{OC}}$: Auxiliary open circuit ``inductor'' in parallel with~$Q_2$. $Q_2$ is simultaneously assumed to be in series with~$L_2$ and in parallel with~$L_2^{\text{OC}}$. Dotted circles: Subnetworks where the~KVL and KCL constraints are imposed. Dashed boxes: Subnetworks associated with the~KVL and KCL equations of motion.}
	\label{fig::maximally:nonlinear:circuit}
\end{figure*}

A circuit reaches its maximum nonlinearity when both the inductive and capacitive subnetworks contain nonlinear elements. Figure~\ref{fig::maximally:nonlinear:circuit}~(a) depicts the simplest configuration of a maximally nonlinear circuit, where the resistively and capacitively shunted junction \emph{parallel} circuit equivalent of a Josephson tunnel junction~(JTJ) is connected to its dual circuit, represented by the equivalent \emph{series} circuit of a quantum phase slip~(QPS).

In this setup, the~JTJ functions as a phase-controlled current source, while the~QPS serves as a charge-controlled voltage source with constitutive relation reading as~\cite{Mooij:2006}
\begin{equation}
	v ( q_2 )
	=
	V_{\textrm{c} 0} \, \sin ( k_{\text{Q}} q_2 ) ,
	\label{eq::QPS:constitutive:relation}
\end{equation}
where~$V_{\textrm{c} 0}$ is the QPS's critical voltage and $k_{\textrm{Q}} = 2 \pi / e$~\footnote{A nontrivial extension could involve nonlinear sources controlled by voltages and currents rather than phases and charges.}. For simplicity, we disregard any fluctuations, which can be readily accounted for by adding parallel current and series voltage noise sources to the JTJ and QPS, respectively.

This circuit must be analyzed by using both~KVL and KCL constraints, leading to a hybrid set of equations of motion comprising both KCL and KVL equations. We can follow either of two approaches, which are dual of each other. Here, we choose to connect an auxiliary open circuit~(OC) in parallel with the~QPS. This~OC can be thought as a very large inductor with inductance~$L_2^{\text{OC}}$, which, in the limit, is infinitely large, $L_2^{\text{OC}} \rightarrow +\infty$. Under these conditions, the current through this auxiliary element is always zero, $\accentset{\bullet}{q}_{L_2^{\text{OC}}}=0$. The dual approach would be to connect a short circuit~(SC) capacitor~$C_1^{\text{SC}}$ in series with the~JTJ.

In order to find the circuit Hamiltonian, we turn off all nonconservative elements by setting all parallel resistors to~OCs and all series resistors to~SCs. In this case, $R_1 \rightarrow \text{OC}$ and $R_2 \rightarrow \text{SC}$. Figure~\ref{fig::maximally:nonlinear:circuit}~(b) shows this new conservative circuit. This method allows us to maintain the proper circuit topology, while eliminating the nonconservative elements, which will be added at the end by means of a hybrid version of the Rayleigh dissipation function.

The conservative circuit comprises~$n=3$ nodes and $b=5$ branches (including that associated with~$L_2^{\text{OC}}$). For a linear circuit, the optimal strategy would be to impose $5-3+1=3$~KVL constrains, leading to $3-1=2$~KCL equations of motion. However, this approach would require to use exclusively phases (or voltages) as independent variables. This would work for the~JTJ but not for the~QPS, which is controlled in charge! Thus, we must resort to a hybrid approach, where, alongside the three~KVL constraints, we also impose one~KCL constraint at node~$\Circled{2}$, KCL~2 (constraints indicated as dotted circles in the figure). With such constraints, the~QPS element~$Q_2$ within the~$L_2-L_2^{\text{OC}}-Q_2$ subnetwork can be simultaneously treated as being both in parallel with~$L_2^{\text{OC}}$ and in series with~$L_2$. This hybrid manifestation of~$Q_2$ is the key to properly study this circuit. The hybrid constraints lead to the~KCL and KVL equations of motion indicated by squares in the figure.

The generalized coordinates and velocities and the corresponding canonical coordinates associated with the two main circuit's subnetworks are: $J_1-C_1 \rightarrow \{ \phi_1 , \accentset{\bullet}{\phi}_1 \} , \{ \phi_1 , q_1 \}$. $L_2^{\text{OC}}-Q_2 \rightarrow \{ \phi_2 , \accentset{\bullet}{\phi}_2 \} , \{ \phi_2 , q_2 \}$. In addition, we must define a set of \emph{hybrid hidden (internal) coordinates} associated with~$L_2$: $\{ \phi , \accentset{\bullet}{q} \}$; these hidden coordinates eventually disappear from the circuit Hamiltonian and from the equations of motion.

Using the sign conventions in the figure, the only nontrivial~KVL constraint, KVL~3, leads to the condition (up to an arbitrary constant set to zero for the phases)
\begin{equation}
	\accentset{\bullet}{\phi} = \accentset{\bullet}{\phi}_1 - \accentset{\bullet}{\phi}_2 \rightarrow \phi = \phi_1 - \phi_2 .
\end{equation}
This condition implies that~$L_2$ acts simultaneously as part of the~$L_2(-L_2^{\text{OC}})-Q_2$ subnetwork as well as the ``coupling'' element between this subnetwork and the parallel~$J_1-C_1$ subnetwork. Without this condition, $J_1-C_1$ and $L_2-Q_2$ would behave as two independent free resonators. The~KCL~2 constraint leads to (up to an arbitrary constant set to zero for the charges)
\begin{equation}
	\accentset{\bullet}{q} = \accentset{\bullet}{q}_2 + \accentset{\bullet}{q}_{L_2^{\text{OC}}} = \accentset{\bullet}{q}_2 \rightarrow q = q_2 ,
\end{equation}
since~$L_2^{\text{OC}}$ is actually an~OC. Again, $Q_2$ is simultaneously in parallel with~$L_2^{\text{OC}}$ and in series with~$L_2$.

The circuit's energy can be written as
\begin{equation}
	\mathcal{E} = \mathcal{E}_1 + \mathcal{E}_2 ,
\end{equation}
where~$\mathcal{E}_1$ is the contribution due to the parallel subnetwork~$J_1-C_1$,
\begin{equation}
	\mathcal{E}_1 = -E_{\text{J}0} \cos \left( k_{\text{J}} \phi_1 \right) + \dfrac{1}{2} C_1 \accentset{\bullet}{\phi}_1^2 ,
\end{equation}
and~$\mathcal{E}_2$ is the contribution due to the series subnetwork~$L_2(-L_2^{\text{OC}})-Q_2$. This second contribution has two equivalent manifestations:
\begin{enumerate}[(1)]
	\item Series constraints:
		\begin{equation}
			\mathcal{E}_2^{\text{s}} = \dfrac{1}{2} L_2 \accentset{\bullet}{q}^2 - E_{\text{Q}0} \cos \left( k_{\text{Q}} q \right) + \dfrac{1}{2} L_2^{\text{OC}} \accentset{\bullet}{q}_{L_2^{\text{OC}}}^2 ,
		\end{equation}
	where~$E_{\text{Q}0} = V_{\text{c}0} / k_{\text{Q}}$ is the~QPS energy.
	\item Hybrid constraints:
		\begin{eqnarray}
			\mathcal{E}_2^{\text{h}} & = & \dfrac{1}{2} L_2 \accentset{\bullet}{q}^2 - E_{\text{Q}0} \cos \left( k_{\text{Q}} q_2 \right) + \dfrac{\phi_2^2}{2 L_2^{\text{OC}}}
			\nonumber\\
			& = & T_2(\accentset{\bullet}{q}) + U_2(q_2 , \phi_2) ,
				\label{eq::E2h}
		\end{eqnarray}
		  where~$T_2$ and $U_2$ are the kinetic and potential energies of the subnetwork written in hybrid form.
\end{enumerate}

The circuit Hamiltonian is found by applying condition~(\ref{eq:generalized:phij}) to Eq.~(\ref{eq::E2h}) and imposing the constraint~KVL~3 to eliminate the hidden coordinate~$q$:
\begin{eqnarray}
	\dfrac{\partial}{\partial \accentset{\bullet}{q}} \mathcal{E}_2^{\text{h}} = \dfrac{\partial}{\partial \accentset{\bullet}{q}} T_2(\accentset{\bullet}{q}) & = & L_2 \accentset{\bullet}{q}
								\nonumber\\
								& = & -\phi = -\left( \phi_1 - \phi_2 \right) ,
\end{eqnarray}
from which it follows that
\begin{equation}
	\accentset{\bullet}{q} = -\dfrac{\phi_1 - \phi_2}{L_2}
		\label{eq::qdot:remove:hidden:coordinate}
\end{equation}
\footnote{Note that, as always, we could have substituted the constraint~$\phi = \phi_1 - \phi_2$ directly into the potential energy~$\phi^2/2 L_2$.}.

The circuit Hamiltonian is obtained by substituting the condition~(\ref{eq::qdot:remove:hidden:coordinate}) into the expression for~$\mathcal{E}$ and transforming the trivial kinetic energy term associated with~$\accentset{\bullet}{\phi}_1$ into the corresponding term in~$q_1$, resulting in
\begin{eqnarray}
	\mathcal{H} & = & -E_{\text{J}0} \cos \left( k_{\text{J}} \phi_1 \right) + \dfrac{q_1^2}{2 C_1}
	\nonumber\\
	& + & \dfrac{\phi_1^2}{2 L_2} - \dfrac{\phi_1 \phi_2}{L_2} + \dfrac{\phi_2^2}{2 L_2}
	\nonumber\\
	& - & E_{\text{Q}0} \cos \left( k_{\text{Q}} q_2 \right) + \dfrac{\phi_2^2}{2 L_2^{\text{OC}}} .
\end{eqnarray}

The circuit dissipation function must be written in a hybrid form, where~$R_1$, belonging to the parallel subnetwork, is associated with~$\accentset{\bullet}{\phi}_1$, and $R_2$, belonging to the series subnetwork, is associated with~$\accentset{\bullet}{q}_2$:
\begin{equation}
	\mathcal{D} = \dfrac{\accentset{\bullet}{\phi}_1^2}{2 R_1} + \dfrac{1}{2} R_2 \accentset{\bullet}{q}_2^2 .
\end{equation}

The first-order equations of motion are found from Eqs.~(\ref{eq:genPoisson:qkdot}) and (\ref{eq:genPoisson:phikdot}) and read as
\begin{subequations}
	\begin{empheq}[left=\empheqlbrace]{align}
		\accentset{\bullet}{q}_1
		& =
		- I_{\text{c}0} \sin \left( k_{\text{J}} \phi_1 \right) - \dfrac{\phi_1 - \phi_2}{L_2} - \dfrac{\accentset{\bullet}{\phi}_1}{R_1}
		\label{eq::q1dot:max:nonlin}
		\\[2mm]
		\accentset{\bullet}{\phi}_1
		& =
		\dfrac{q_1}{C_1}
		\label{eq::phi1dot:maxnonlin}
		\\[2mm]
		\accentset{\bullet}{q}_2
		& =
		\dfrac{\phi_1 - \phi_2}{L_2} - \dfrac{\phi_2}{L_2^{\text{OC}}}
		\label{eq::d2dot:maxnonlin}
		\\[2mm]
		\accentset{\bullet}{\phi}_2
		& = V_{\text{c}0} \sin \left( k_{\text{Q}} q_2 \right) + R_2 \accentset{\bullet}{q}_2 .
		\label{eq::phi2dot:maxnonlin}
	\end{empheq}
\end{subequations}
Since~$L_2^{\text{OC}}$ is an~OC, Eq.~(\ref{eq::d2dot:maxnonlin}) can be written as
\begin{equation}
	\Lim{L_2^{\text{OC}} \rightarrow +\infty} \accentset{\bullet}{q}_2 = \dfrac{\phi_1 - \phi_2}{L_2} ,
		\label{eq::lim:q2dot:maxnonlin}
\end{equation}
which is also one of the three current contributions in Eq.~(\ref{eq::q1dot:max:nonlin}). From~Eq.~(\ref{eq::lim:q2dot:maxnonlin}) follows that
\begin{equation}
	\phi_2 = \phi_1 - L_2 \accentset{\bullet}{q}_2 .
\end{equation}

The four first-order equations of motion can be then rewritten as two second-order equations of motion,
\begin{subequations}
	\begin{empheq}[left=\empheqlbrace]{align}
		& \dfrac{\accentset{\bullet}{\phi}_1}{R_1} + I_{\text{c}0} \sin \left( k_{\text{J}} \phi_1 \right) + C_1 \accentset{\bullet\bullet}{\phi}_1
		\nonumber\\
		& + \accentset{\bullet}{q}_2 = 0
		\label{eq::KCL1:max:nonlin}
		\\[2mm]
		& -\accentset{\bullet}{\phi}_1
		\nonumber\\
		& + L_2 \accentset{\bullet\bullet}{q}_2 + V_{\text{c}0} \sin \left( k_{\text{Q}} q_2 \right) + R_2 \accentset{\bullet}{q}_2 = 0 .
		\label{eq::KVL3:maxnonlin}
	\end{empheq}
\end{subequations}
which coincide with~KCL~1 and KVL~3 as obtained from direct circuit inspection.

\section{Conclusion}

In conclusion, we introduce a general method that, from a circuit's digraph, allows us to identify whether to use KCLs or KVLs. By integrating the circuit's power we find the energy, and by suitably transforming the energy the Hamiltonian. We show the necessity of auxiliary circuit elements in presence of incomplete sets of canonical coordinates and two distinct procedures to eliminate them: Reduction at the Hamiltonian or dissipation function level or after obtaining the equations of motion. We outline a general method to account for noisy resistors. We present the quantum version of the previously introduced classical methods, including dissipation and fluctuations. Finally, we consider maximally nonlinear and nonconservative circuits with both~JTJs and QPSs, leading to a hybrid formalism that uses both~KVL and KCL constraints to obtain~KCL and KVL mixed equations of motion.

\begin{acknowledgements}
We would like to thank useful discussions with N.~Gorgichuk.
\end{acknowledgements}

\bibliography{Bibliography}

\end{document}